\documentclass[
]{ceurart}

\usepackage{listings}
\usepackage{graphicx}
\usepackage{float}
\usepackage{bbding}
\usepackage{enumerate}
\usepackage{color,soul}
\usepackage{graphicx}
\usepackage{float}
\usepackage{enumitem}
\usepackage{bbding}
\usepackage{amsmath}
\usepackage{xcolor}
\usepackage{colortbl}
\usepackage{fancybox}
\usepackage{amssymb}
\usepackage{enumitem}
\usepackage{pdfpages}
\usepackage{tikz}
\usepackage{array}
\usepackage{multirow}
\usepackage{capt-of}
\usepackage{color,soul}
\usepackage{longtable}
\usepackage{forest}
\usepackage{multirow}
\usepackage{tabularx}
\usepackage{eqnarray}
\usepackage{color,soul}
\usepackage{ltablex} 
\usepackage{ltablex}
\usepackage{supertabular}
\usepackage{arydshln}
\usepackage[linesnumbered,ruled]{algorithm2e}
\usepackage{hyperref}
\usepackage{url}
\usepackage[latin1]{inputenc}
\usepackage{makecell}
\usepackage{listings}
\usepackage{xcolor}
\usepackage{array}
\usepackage{multirow}
\usepackage{ragged2e}
 
\definecolor{codegreen}{rgb}{0,0.6,0}
\definecolor{codegray}{rgb}{0.5,0.5,0.5}
\definecolor{codepurple}{rgb}{0.58,0,0.82}
\definecolor{backcolour}{rgb}{0.95,0.95,0.92}
\lstdefinestyle{mystyle}{
    backgroundcolor=\color{backcolour},   
    commentstyle=\color{codegreen},
    keywordstyle=\color{magenta},
    numberstyle=\tiny\color{codegray},
    stringstyle=\color{codepurple},
    basicstyle=\ttfamily\footnotesize,
    breakatwhitespace=false,         
    breaklines=true,                 
    captionpos=b,                    
    keepspaces=true,                 
    numbers=left,                    
    numbersep=5pt,                  
    showspaces=false,                
    showstringspaces=false,
    showtabs=false,                  
    tabsize=2
}
 
\begin{document}
\copyrightyear{2023}
\copyrightclause{Copyright for this paper by its authors.
  Use permitted under Creative Commons License Attribution 4.0
  International (CC BY 4.0).}

\conference{Forum for Information Retrieval Evaluation (FIRE)- 2023, 
Indian Statistical Institute, Kolkata, India, 
$15^{th}-13^{th}$ December, 2023  }

\title{Generative AI for Software Metadata: Overview of the Information Retrieval in Software Engineering Track at FIRE 2023}

%tnotemark[1]
%\tnotetext[1]{You can use this document as the template for preparing your
%  publication. We recommend using the latest version of the ceurart style.}
                                                                                                             
\author[1,2]{Srijoni Majumdar}[%
 email=majumdar.srijoni@gmail.com,
]
\cormark[1]
\fnmark[1]
\author[1]{Soumen Paul}[%
 email=soumenpaul165@gmail.com,
]
\cormark[1]
\fnmark[1]

\author[8]{Debjyoti Paul}[%
email=debjyoti93.paul@gmail.com ,
]

\author[4]{Ayan Bandyopadhyay}[%
email=bandyopadhyay.ayan@gmail.com,
]

\author[7]{Samiran Chattopadhyay}[%
email=samiran.chattopadhyay@jadavpuruniversity.in, 
]

\author[1]{Partha Pratim Das}[%
email=ppd@cse.iitkgp.ac.in,
]

\author[4,5]{Paul D Clough}[%
email=p.d.clough@sheffield.ac.uk,
]

\author[3, 6]{Prasenjit Majumder}[%
email=prasenjit.majumder@gmail.com,
]

\address[1]{IIT Kharagpur, West-Bengal, India}
\address[2]{University of Leeds,  UK}
\address[3]{TCG CREST, West-Bengal, India}

\address[4]{TPXimpact London, UK}
\address[5]{Sheffield University, Sheffield, UK}
\address[6]{DA-IICT Gandhinagar, Gujarat, India}
\address[7]{Jadavpur University, West-Bengal, India}
\address[8]{Indian Statistical Institute, Kolkata India}
\cortext[1]{Corresponding author.}
\fntext[1]{These authors contributed equally.}

\begin{abstract}
The Information Retrieval in Software Engineering (IRSE) track aims to develop solutions for automated evaluation of code comments in a machine learning framework based on human and large language model generated labels. In this track, there is a binary classification task to classify comments  as useful and not useful. The dataset consists of 9048 code comments and surrounding code snippet pairs extracted from open source {\em github} C based projects and an additional dataset generated individually by teams using large language models. Overall 56 experiments have been submitted by 17 teams from various universities and software companies. The submissions have been evaluated quantitatively using the F1-Score and qualitatively based on the type of features developed, the supervised learning model used and their corresponding hyper-parameters.  The labels generated from large language models increase the bias in the prediction model but lead to less over-fitted results.
\end{abstract}

\begin{keywords}
bert \sep GPT-2 \sep Stanford POS Tagging \sep neural networks \sep abstract syntax tree
\end{keywords}

\maketitle

\section{Introduction}
Assessing comment quality can help to de-clutter code bases and subsequently improve code maintainability. 
Comments can significantly help to read and comprehend code if they are consistent and informative. 

The perception of quality in terms of the 'usefulness' of the information contained in comments is relative and hence is perceived differently based on the context. Bosu et al.~\citep{bosu2015characteristics} attempted to assess code review comments  (logged in a separate tool) in the context of their utility in helping developers write better code through a detailed survey at Microsoft. A similar quality assessment model is important to analyse the type of source code comments that can help for standard maintenance tasks but is largely missing. Majumdar et al.~\cite{majumdar2022automated} proposed a comment quality evaluation framework wherein comments were assessed as 'useful', 'partially useful', and 'not useful'  based on whether they increase the readability of the surrounding code snippets. The authors analyse comments for concepts that aid in code comprehension and also the redundancies or inconsistencies of these concepts with the related code constructs in a machine learning framework for an overall assessment. The concepts are derived through exploratory studies with developers across 7 companies and from a larger community using crowd-sourcing.

The first edition of the IRSE track of FIRE 2023, extends the work in~\cite{majumdar2022automated} and  empirically investigates comment quality with a larger set of machine learning solvers and  features.  The task is based on  the quality evaluation of comments into two clusters - 'useful' and 'not useful'. A 'useful' comment (refer Table~\ref{tab:data}) contains relevant concepts that are not evident from the surrounding code design, and thus increases the comprehensibility of the code. The suitability of analysing comment quality using various vector space representations of code and comment pairs along with standard textual features and code comment correlation links are evaluated. 

\begin{table}[!ht]
	\caption{Useful and Not-Useful comments in context of code comprehension}
	\centering
	\begin{footnotesize}
		\begin{tabular}{|p{0.1cm}|p{2.3cm}|p{8cm}|p{0.8cm}|}
			\hline
			 %\rowcolor{lightgray}
		 \multicolumn{1}{|p{0.1cm}|}{\bf \#} &		 \multicolumn{1}{|p{2.3cm}|}{\bf Comment}   & 			\multicolumn{1}{|p{6cm}|}{\bf Code} & \multicolumn{1}{|p{0.8 cm}|}{ \bf Label}  \\ \hline

1 & /* uses png\_calloc defined 
in pngriv.h*/

& 
{\em /* uses png\_calloc defined in pngriv.h*/}

\begin{minipage}[t]{0.2\textwidth}
\begin{verbatim}
PNG_FUNCTION(png_const_structrp png_ptr)

{  if (png_ptr == NULL || info_ptr == NULL)
      return;
   png_calloc(png_ptr); ...}
   \end{verbatim}
\end{minipage} & U\\ 

2 & /* serial bus is locked before use */

& 
\begin{minipage}[t]{0.2\textwidth}
\begin{verbatim}
static int bus_reset ( . . . ) /* serial bus 
is locked before use*/
{ .. update_serial_bus_lock (bus * busR); }
   \end{verbatim}
\end{minipage} & NU\\

3 & // integer variable 

& \begin{minipage}[t]{0.2\textwidth}
\begin{verbatim}
int Delete\_Vendor;  // integer variable 
   \end{verbatim}
\end{minipage}

& NU \\
 \hline
\multicolumn{4}{c}{U: Useful; NU: Not Useful}			
			\end{tabular}	
		\label{tab:data} 
	\end{footnotesize}
\end{table}

The 2023 IRSE track extends this challenge to understand the feasibility of using silver standard quality labels generated from the Large Language Models (LLMs) and understand how it augments the classification model in terms of prediction. Developing the gold industry standard for analysing the usefulness of comments that can be relevant for code comprehension in legacy systems can be challenging and time-consuming. However, to scale the model and use it on different languages, it is important to generate more data which we attempt to do with the large language models. The performance of these modes in the context of understating the relations between code and comment can provide an approximation of the data quality generated and how it can be used to scale the existing classification mode. This approach can also be further generalised to any classification model based on software metadata.

\section{Related Work}

Software metadata is integral to code maintenance and subsequent comprehension. A significant number of tools~\cite{majumdar2019smartkt, chatterjee2015debugging,majumdar2016d,majumdar2021mathematical,majumdar2021dcube_,o2003software}  have been proposed to aid in extracting knowledge from software metadata like runtime traces or structural attributes of codes.

In terms of mining code comments and assessing the quality, authors~\cite{steidl2013quality,majumdar2022can,majumdar2022overview} compare the similarity of words in code-comment pairs using the Levenshtein distance and length of comments to filter out trivial and non-informative comments. Rahman et al.~\citep{rahman2017predicting} detect useful and non-useful code review comments (logged-in review portals) based on attributes identified from a survey conducted with developers of Microsoft~\citep{bosu2015characteristics}. Majumdar et al.~\cite{majumdar2022automated,majumdar2020comment} proposed a framework to evaluate comments based on concepts that are relevant for code comprehension. They developed textual and code correlation features using a knowledge graph for semantic interpretation of information contained in comments.
These approaches use semantic and structural features to design features to set up a prediction problem for useful and not useful comments that can be subsequently integrated into the process of decluttering codebases.

With the advent of large language models~\cite{brown2020language}, it is important to compare the quality assessment of code comments by the standard models like GPT 3.5 or llama with the human interpretation.  The IRSE  track at FIRE 2023 extends the approach proposed in~\cite{majumdar2022automated} to explore various vector space models~\cite{majumdar2022effective} and features for binary classification and evaluation of comments in the context of their use in understanding the code. This track also compares the performance of the prediction model with the inclusion of the GPT-generated labels for the quality of code and comment snippets extracted from open-source software.

\section{IRSE Track Overview and Data Set}

The following section outlines the task descriptions and the characteristics of the dataset.

\subsection{Task Description}

{\em Comment Classification}: A binary classification task to classify source code comments as {\em Useful} or {\em Not Useful} for a given comment and associated code pair as input.

\noindent {\tt Input}: A code comment with surrounding code snippet (written in C)

\noindent {\tt Output}: A label (Useful or Not Useful) that characterises whether the comment helps developers comprehend the associated code

Therefore in this classification task, the output is based on whether the information contained in the comment is relevant \emph{and} would help to comprehend the surrounding code, i.e., it is \emph{useful}.

{\bf Useful}: Comments have sufficient software development concept $\rightarrow$ Comment is Relevant, and these concepts are not mostly present in the surrounding code $\rightarrow$ Comment is not Redundant, hence the comment is {\em Useful}

{\bf Not Useful}: Comments have sufficient software development concept $\rightarrow$ Comment is Relevant, and these concepts are  mostly present in the surrounding code $\rightarrow$ Comment is  Redundant, hence the comment is {\em Not Useful}

It may also be the case that comments do not contain sufficient software development concepts $\rightarrow$ Comment is Not Relevant, hence the comment is {\em Not Useful}. 

It is left to the participants to decide on the threshold value for how many concepts retrieved make a comment relevant or how many matches with surrounding code make a comment redundant.

The notion of {\em relevant} comments refers to those that developers perceive as important in comprehending the associated or surrounding lines of code. These concepts are related to the outline of the algorithm, data-structure descriptions, mapping to user interface details, possible exceptions, version details, etc. In the below examples, the comments highlight useful details about the input data to the function, which is not evident from the associated code itself.

{\bf Dataset}: For the IRSE track, we use  a set of 9048 comments (from Github) with comment text, surrounding code snippets, and a label that specifies whether the comment is useful or not. Sample data has been characterised in Table~\ref{tab:data}.

\begin{itemize}
\item The development dataset contains 8048 rows of comment text, surrounding code snippets, and labels (Useful and Not useful). 

\item The test dataset contains 1,000 rows of comment text, surrounding code snippets, and labels (Useful and Not useful). 

\end{itemize}

\begin{table}
\centering
\caption{Characterizations of the Submissions: test Data Predictions}
\begin{tabular}{|c|l|l|l|l|l|l|l|l|} 
\hline
\multirow{2}{*}{\begin{tabular}[c]{@{}c@{}}Affiliation  \end{tabular}} & \multicolumn{3}{c|}{Seed data}        & \multicolumn{3}{c|}{Seed data + LLM-generated data}  \\ 
\cline{2-9}
                                                                     & Precision & Recall & F1Score    & Precision & Recall & F1Score                   \\ 
\hline
\multirow{3}{*}{DSTI, France}                                                & 0.8326    & 0.8626 & 0.8473   & 0.844     & 0.8682 & 0.8559                 \\ 
\cline{2-9}
                                                                     & 0.8948    & 0.8738 & 0.884   &  0.9       & 0.8707 & 0.885                \\ 
\cline{2-9}
                                                                     & 0.8807    & 0.8822 & 0.8813  &  0.8871    & 0.8839 & 0.8854             \\ 
\hline
\multirow{2}{*}{SSN-1 (RAM)}                                                 & 0.8       & 0.8    & 0.8     & 0.8021 & 0.81      & 0.73                      \\ 
\cline{2-9}
                                                                     & 0.72      & 0.71   & 0.74   & 0.7       & 0.73   & 0.74                        \\ 
\hline
\multirow{5}{*}{SSN-2 (Aloy)}                                                & 0.788     & 0.7363 & 0.7613        & 0.89      & 0.8802 & 0.8846                       \\ 
\cline{2-9}
                                                                     & 0.7994    & 0.7994 & 0.7994        & 0.89      & 0.8795 & 0.8841                        \\ 
\cline{2-9}
                                                                     & 0.7993    & 0.9352 & 0.8619         & 0.839     & 0.9199 & 0.8776                        \\ 
\cline{2-9}
                                                                     & 0.7842    & 0.8453 & 0.8136        & 0.8154    & 0.8823 & 0.8475                       \\ 
\cline{2-9}
                                                                     & 0.7572    & 0.8637 & 0.807          & 0.7785    & 0.9003 & 0.835                          \\ 
\hline
\begin{tabular}[c]{@{}c@{}}IIT (ISM)\\Dhanbad\end{tabular}           & 0.92      & 0.96   & 0.94   & 0.92      & 0.97   & 0.97                  \\ 
\hline
\multirow{4}{*}{SSN-3 (Black)}                                               & 0.7916    & 0.8446 & 0.8172  &     0.7886    & 0.847  & 0.8167                         \\ 
\cline{2-9}
                                                                     & 0.763     & 0.8696 & 0.813          & 0.7655    & 0.8724 & 0.8154                         \\ 
\cline{2-9}
                                                                     & 0.705     & 0.9387 & 0.8052         & 0.6994    & 0.9041 & 0.7887                         \\ 
\cline{2-9}
                                                                     & 0.7292    & 0.856  & 0.7875          & 0.7374    & 0.8533 & 0.7911                        \\ 
\hline
\begin{tabular}[c]{@{}c@{}}Microsoft-\\American Express\end{tabular}             & 0.7902    & 0.8016 & 0.7949  & 0.7908    & 0.8014 & 0.7952                 \\ 
\hline
DDU-1                                                            &   0.895        &     0.891   &       0.893    &       0.890    &    0.894    &   0.892                     \\ 
\hline
DDU-2                                                            &   0.875        &     0.872  &       0.874  &       0.870    &    0.875    &   0.880                        \\  
\hline
IIT KGP-1                                                             & 0.8283    & 0.804  & 0.8141   & 0.8322    & 0.8086 & 0.8185               \\ 
\hline
SRM                                                              & 0.8283    & 0.804  & 0.8141  & 0.8178    & 0.7906 & 0.8013                  \\ 
\hline
IIT KGP-2                                                              & 0.78      & 0.85   & 0.8   & 0.77      & 0.85   & 0.8                \\ 
\hline
DA-IICT                                                             & 0.81      & 0.8    & 0.8           & 0.58      & 0.58   & 0.58                         \\ 
\hline
IIT Goa                                                                & 0.6087    & 0.6526 & 0.6321         & 0.6114    & 0.6598 & 0.6403                      \\ 
\hline
\begin{tabular}[c]{@{}c@{}}TCS\end{tabular}                 & 0.778    & 0.753 & 0.74   & 0.645    & 0.6598 & 0.650              \\
\hline

IIT KGP-3                                                              & 0.631    & 0.645 & 0.639         & 0.6114    & 0.6598 & 0.631                        \\ 
\hline

Amazon                                                               & 0.659    & 0.672 & 0.666          & 0.656    & 0.635 & 0.645                         \\ 
\hline
\end{tabular}
	\label{tab:static1} 
\end{table}

\begin{table}
\centering
\caption{Characterizations of the LLM Generated datasets}
\begin{tabular}{|l|l|l|l|} 
\hline
Team name         & Total entry & Useful entry & Not useful entry  \\ 
\hline
DSTI, France             & 421         & 412          & 9                 \\ 
\hline
SSN-1 (RAM)               & 1238        & 740          & 497               \\ 
\hline
SSN-2 (Alloy)             & 1510        & 24           & 1486              \\ 
\hline
IIT (ISM) Dhanbad & 199         & 182          & 17                \\ 
\hline
SSN 3 (Black)            & 738         & 80           & 658               \\ 
\hline
Microsoft - American Express   & 233         & 92           & 141               \\ 
\hline
DDU-1         & 8588        & 4649         & 3939              \\ 
\hline
DDU-2          & 332         & 311          & 21                \\ 
\hline
IIT KGP-1          & 334         & 309          & 25                \\ 
\hline
SRM         & 217         & 196          & 21                \\ 
\hline
IITKGP-2           & 263         & 130          & 133               \\ \hline
DA-IICT          & 150        & 65          & 85           \\ 
\hline
IIT-Goa            & 543         & 460          & 83                \\ 
\hline

TCS        & 282         & 61           & 221               \\\hline
IITKGP-3            & 570        & 450          & 120                \\ \hline
IITKGP-3            & 412        & 345         & 67                \\ 
\hline

\end{tabular}
\label{tab:static2} 

\end{table}

\section{Participation and Evaluation}

IRSE 2023 received a total of 56 experiments from 17 teams for the two tasks. As this track is related to software maintenance,  we received participation from  companies like Microsoft, Amazon, Amercian  Express, Bosch Research along with several research labs of educational institutes.

The various teams with the details of their submissions are characterised in Table~\ref{tab:static1}.

{\em Evaluation Procedure}: Candidates submited the prediction metrics (precision,recall, F1-Score) fo the classification model with the Gold labels dataset (referred to as the Seed Dataset) and combined dataset (Seed + LLM generated labels - Silver lables dataset). The difference in the F1 score was evaluated by us.

{\em Features}: Apart from evaluating the prediction metrics, we analysed the types of features the teams have used to devise the machine learning pipeline. The teams have performed routine pre-processing and have retained the significant words or letters only for both the code and comment pairs. Further, some of the teams have also used morphological features of a comment like a length, significant words ratio, parts of speech characteristics, or occurrence of words from an enumerated set as textual features. To correlate code and comment and detect redundancies, the teams mostly used grep-like string match to find similar words. 

{\em Vector Space Representations}: Code and comments belong- to different semantic granularity which is unified by a vector space representation. The participants have used various pre-trained embeddings to generate vectors for the words like those based on one hot encoding, tf-idf based, word2vec or context aware like ELMo and BERT. Each of the employed embedding models are trained or finetuned using software development corpora.

{\em Results}: The participants are able to achieve a slight increase (in the range of 2\%-4\%) in the test prediction metrics and in many cases the performance decrease. However, the increase in bias due to the incorporation of silver standard data reduces the over-fitting of the models.

% \begin{figure}[!ht]
% \begin{center}
% 	\includegraphics[scale=0.36]{model.png} 
% 		\caption{Distribution of Machine Learning Architectures - 34 experiments}
% 		\label{fig:archi1}
% 	\end{center}
% \end{figure}

%%

% \begin{figure}[!ht]
% \begin{center}
% 	\includegraphics[scale=0.36]{pre_trained.png} 
% 		\caption{Distribution of Pre-trained embeddings}
% 		\label{fig:archi2}
% 	\end{center}
% \end{figure}

\section{Conclusions}

The IRSE 2023 track empirically investigates the feasibility of augmenting existing classification models using datasets with labels generated from LLM's. A total of 17 teams participated and submitted 56 experiments that used various types of machine learning models, embedding spaces, features and different LLMs to generate data. The LLM-generated labels reduce the overfitting of the overall classification model and also improve the F1 score when the combined data from all participants were used to augment the existing data with gold standard labels from industry practitioners.

\bibliography{sample-ceur}
%\bibliography{sample-base}

\end{document}